\documentclass[11pt]{article}
\usepackage[dvips]{graphicx,psfrag}

\begin{document}
% ------------------------------------------------------------------------ %
\title{Exact Formula of Probability and CP Violation\\
for Neutrino Oscillations in Matter}
\author{
{K. Kimura$^1$}\thanks{E-mail address:kimukei@eken.phys.nagoya-u.ac.jp} 
{, A. Takamura$^{1,2}$}\thanks
{E-mail address:takamura@eken.phys.nagoya-u.ac.jp} 
{, and H. Yokomakura$^1$}\thanks{E-mail address:yoko@eken.phys.nagoya-u.ac.jp} 
\\
\\
\\
{\small \it $^1$Department of Physics, Nagoya University,}
{\small \it Nagoya, 464-8602, Japan}\\
{\small \it $^2$Department of Mathematics, 
Toyota National Collage of Technology}\\
{\small \it Eisei-cho 2-1, Toyota-shi, 471-8525, Japan}}
\date{}
\maketitle
% ------------------------------------------------------------------------ %
\vspace{-11cm}
\begin{flushright}
%hep-ph/0005075\\
%DPNU-02-05\\
%KEK-TH-692\\
\end{flushright}
\vspace{10.5cm}
\vspace{-2.5cm}
% ======================================================================== %
%
% abstract
%
% ------------------------------------------------------------------------ %
\vspace{1cm}
\begin{abstract}
Within the framework of the standard three neutrino scenario, 
we derive an exact and simple formula of 
the oscillation probability 
$P(\nu_e \to \nu_{\mu})$ in constant matter by using a new method.
From this formula, it is found that 
the matter effects can be separated from the pure 
CP violation effects.
Furthermore, the oscillation probability can be written in the form, 
$P(\nu_e \to \nu_{\mu})=A\cos \delta+B\sin\delta+C$, 
in the standard parametrization without any approximation.
We also demonstrate that the approximate formula in high-energy 
can be easily reproduced from this as an example.
\end{abstract}

%{\sf PACS:12.15.Ff, 14.60.Pq, 23.40.Bw.}%

\section{Introduction}
\hspace*{\parindent}
Just like the quark system, 
it has been shown from the atmospheric neutrino experiments 
\cite{atmospheric} and the solar neutrino experiments 
\cite{solar} that neutrinos have finite mass and 
finite mixing.
In this situation, it is extremely interesting to investigate 
the CP phase in the lepton sector.
Fortunately, recent report from SNO experiment \cite{SNO} 
favors the LMA MSW solutions to the solar neutrino problem.
This means that the measurements of CP phase may be 
possible because of the large 1-2 mixing angle and the 
large 1-2 mass difference.

In order to measure the CP phase, 
the long-baseline experiments such as 
the JHF experiment \cite{JHF} and 
the neutrino factory experiments \cite{nu-factory}
are planned.
In the past, the asymmetries 
$\Delta P_{CP} =P(\nu_{\alpha} \to \nu_{\beta})-
P(\bar{\nu}_{\alpha} \to \bar{\nu}_{\beta})$ and 
$\Delta P_T =P(\nu_{\alpha} \to \nu_{\beta})
-P(\nu_{\beta} \to \nu_{\alpha})$ have been considered 
as the main approach to measure the CP phase $\delta$
\cite{Harrison, Sato, Parke, other1}. 
These are methods to measure the direct CP violation term 
which depends on $\sin \delta$. 
However, the measurement of $\Delta P_{CP}$ 
is not directly related to 
the discovery of CP phase, because of 
fake CP violating effects from the earth matter.
$\Delta P_T$ is a pure T violating observable, but 
it has its own experimental difficulties.
So, alternative approach has been recently considered in  
\cite{Cervera, Freund, Minakata, Lipari, Freund2}.
This is an attempt to obtain the information on 
the CP phase totally from the probabilities itself, 
not only the direct CP violation term but also 
the indirect CP violation term which depends on $\cos \delta$.  
In these papers the oscillation probability is written approximately
in the form, 
$P(\nu_e \to \nu_{\mu})\simeq A\cos\delta+B\sin\delta+C$. 
The extra information which is proportional to $\cos \delta$ 
will lead us to the value of $\delta$ in spite of the matter effect.
In order to obtain more precise information, 
it is highly desirable to have an exact expression for 
$P(\nu_e \to \nu_{\mu})$.
Some attempts to derive the exact formula have been made 
in the context of three neutrino scenarios 
\cite{Barger1, Zaglauer, Xing, Ohlsson}.
These formulae are useful for numerical calculation. 
However, the precise CP dependence of $P(\nu_e \to \nu_{\mu})$ 
has not been investigated  
sufficiently \cite{Zaglauer}.

To describe our approach, let us review the work of 
Naumov \cite{Naumov} and  
Harrison-Scott \cite{Harrison}.
The Hamiltonian $\tilde{H}$ in matter 
is related to $H$ in vacuum as 
\begin{eqnarray}
\tilde{H}=H+\frac{1}{2E}{\rm diag}(a, 0, 0),
\end{eqnarray}
where $a\equiv 2\sqrt{2}G_F N_e E$, $G_F$ is Fermi constant 
and $N_e$ is the electron density in matter.
In particular, taking the products of non-diagonal elements, 
\begin{eqnarray}
{\rm Im}(\tilde{H}_{e\mu}\tilde{H}_{\mu\tau}\tilde{H}_{\tau e})
={\rm Im}(H_{e\mu}H_{\mu\tau}H_{\tau e}),
\end{eqnarray}
one obtains the following identity, which we call 
Naumov-Harrison-Scott identity, 
\begin{eqnarray}
\tilde{\Delta}_{12}\tilde{\Delta}_{23}\tilde{\Delta}_{31}
\tilde{J}=\Delta_{12}\Delta_{23}\Delta_{31}J,
\end{eqnarray}
in CP-odd part, where 
$\Delta_{ij} \equiv m_i^2-m_j^2$,
$J\equiv {\rm Im} J_{e\mu}^{12}$ is Jarlskog factor 
\cite{Jarlskog}, $J_{\alpha \beta}^{ij}\equiv
U_{\alpha i}U_{\beta i}^* U_{\alpha j}^* U_{\beta j}$ and 
$U$ is the Maki-Nakagawa-Sakata (MNS) matrix \cite{MNS}. 
Here the quantities expressed by the tilde include the matter effects.
From this identity, $\tilde{J}$ can be expressed in terms of 
effective masses and the parameters of the Hamiltonian in vacuum.
As effective masses shown in \cite{Barger1, Zaglauer, Xing}
do not depend on the CP phase, 
$\tilde{J}$ can be completely expressed by the linear term in 
$\sin \delta$.
The reason why the CP dependence becomes simple is 
that $\tilde{J}$ is the products of four $\tilde{U}$'s.
Complicated matter effects included in 
a $\tilde{U}$ are partially canceled in $\tilde{J}$.
\footnote{The calculation of a $\tilde{U}$ is performed 
by diagonalizing $\tilde{H}$ in Ref. \cite{Xing}.}

In this letter, we calculate $\tilde{U}\tilde{U}^*$. 
We use some matter invariant identities  
such as Naumov-Harrison-Scott identity
and express not only $\tilde{J}$ but also 
${\rm Re}\tilde{J}_{e\mu}^{ij}$ 
with the effective masses and the parameters in vacuum.
The exact formula obtained in this method 
is very simple and the matter effects come in only through  
effective masses.
We show that 
${\rm Re}\tilde{J}_{e\mu}^{ij}$ has only 
a linear term in $\cos \delta$.
That is, we prove that the oscillation probability in matter 
can be written in the form 
\begin{eqnarray}
P(\nu_e \to \nu_{\mu})=A\cos\delta+B\sin\delta+C
\end{eqnarray} 
without any approximation. 

Another merit of our result 
is that the exact formula immediately reduces to the well-known 
approximate formulae both in low-energy \cite{Sato} and 
in high-energy \cite{Cervera, Freund}.
We demonstrate that the approximate formula in high-energy 
can be easily reproduced from our exact formula as an example.
Finally, we numerically calculate
the coefficients $A$, $B$ and $C$.

\section{Exact Formula of the Oscillation Probability}
\hspace*{\parindent}
The flavor and mass eigenstates are related by the MNS matrix 
$\tilde{U}_{\alpha i}$ in matter, where $\alpha=e, \mu, \tau$ is 
the flavor index, $i=1, 2, 3$ is the mass index.  
The amplitude for $\nu_e$ to $\nu_{\mu}$ transition is given by
\begin{eqnarray}
A(\nu_e \to \nu_{\mu})
=\sum_{i=1}^3
\tilde{U}_{ei}^* e^{-i\frac{\lambda_i}{2E}L}\tilde{U}_{\mu i},
\label{1}
\end{eqnarray}
and the oscillation probability is also given by 
\begin{eqnarray}
P(\nu_e \to \nu_{\mu})=|A(\nu_e \to \nu_{\mu})|^2 
 \label{22},
\end{eqnarray}
from the amplitude, where $L$ stands for the baseline length. 

We note from (\ref{1}) that the amplitude 
depends only on the products 
$\tilde{U}_{e i}\tilde{U}_{\mu i}^*$. 
One of the important points in this letter is that 
these products can be easily calculated from identities 
which we derive below. 

From the unitarity relation and the other two 
relations,  
\begin{eqnarray}
\tilde{H}_{e\mu} &=& H_{e\mu}=p/(2E),
\label{30} \\
\tilde{H}_{e \tau}\tilde{H}_{\tau\mu}
-\tilde{H}_{e\mu}\tilde{H}_{\tau\tau} &=&
H_{e\tau}H_{\tau\mu}-H_{e\mu}H_{\tau\tau}
=q/(2E)^2,  \label{2}
\end{eqnarray}
three identities on the products 
$\tilde{U}_{ei}\tilde{U}_{\mu i}^*$ can be obtained as follows:  
\begin{eqnarray}
\sum_{i=1}^3\tilde{U}_{ei}\tilde{U}_{\mu i}^*
&=&\sum_{i=1}^3 U_{ei} U_{\mu i}^*=0,
\label{3} \\
\sum_{i=1}^3 \lambda_i \tilde{U}_{ei}\tilde{U}_{\mu i}^*
&=&\sum_{i=1}^3 m_i^2 U_{ei}U_{\mu i}^*=p,
\label{4} \\
\sum_{(ijk)}^{{\rm cyclic}}
\lambda_j \lambda_k \tilde{U}_{ei}\tilde{U}_{\mu i}^*
&=&\sum_{(ijk)}^{{\rm cyclic}}
m_j^2 m_k^2 U_{ei}U_{\mu i}^*=q,
\label{5}
\end{eqnarray}
where $p$ and $q$ are constants 
determined by the parameters in vacuum and 
the sum is over $(ijk)=(123), (231), (312)$.
We use the relation 
$\tilde{U}_{\tau i}\tilde{U}_{ej}
-\tilde{U}_{ei}\tilde{U}_{\tau j}
=\tilde{U}_{\mu k}^* ({\rm det} \tilde{U})$ etc, 
obtained from the formula 
$\tilde{U}^{\dagger}=\tilde{U}^{-1}=
\tilde{\cal{U}}({\rm det} \tilde{U})^{-1}$,
where $\tilde{\cal{U}}$ represents the cofactor matrix.

Solving the simultaneous equations for the products 
$\tilde{U}_{ei}\tilde{U}_{\mu i}^*$, we obtain  
\begin{eqnarray}
\tilde{U}_{ei}\tilde{U}_{\mu i}^*=\frac{p\lambda_i+q}
{\tilde{\Delta}_{ji}\tilde{\Delta}_{ki}} \label{206},
\end{eqnarray}
where $(ijk)$ takes $(123),(231),(312)$.
From the definition $\tilde{J}_{e\mu}^{ij}
=\tilde{U}_{ei}\tilde{U}_{\mu i}^* 
(\tilde{U}_{ej}\tilde{U}_{\mu j}^*)^*$, 
the exact formula of the oscillation probability is given by 
\begin{eqnarray}
P(\nu_e \to \nu_{\mu})
=-4\sum_{(ij)}^{{\rm cyclic}}{\rm Re}\tilde{J}_{e\mu}^{ij}
\sin^2 \left(\frac{\tilde{\Delta}_{ij}L}{4E}\right)
- 2\sum_{(ij)}^{{\rm cyclic}}\tilde{J}
\sin \left(\frac{\tilde{\Delta}_{ij}L}{2E}\right) \label{200},
\end{eqnarray}
where the sum is over $(ij)=(12),(23),(31)$ and  
\begin{eqnarray}
{\rm Re}\tilde{J}_{e\mu}^{ij}&=&
\frac{|p|^2 \lambda_i \lambda_j+|q|^2
+{\rm Re}(pq^*)(\lambda_i+\lambda_j)}
{\tilde{\Delta}_{ij}\tilde{\Delta}_{12}\tilde{\Delta}_{23}
\tilde{\Delta}_{31}}, \label{6} \\
\tilde{J}&=&\frac{{\rm Im}(pq^*)}
{\tilde{\Delta}_{12}\tilde{\Delta}_{23}
\tilde{\Delta}_{31}}. \label{7}
\end{eqnarray}
We find that the matter effects are confined in  
the effective masses only.
We can obtain the probability for antineutrinos,  
$\bar{\nu}_e \to \bar{\nu}_{\mu}$, 
by exchanging $a \to -a$ and $\delta \to -\delta$ 
in $\tilde{\Delta}_{ij}$ and $\tilde{J}_{e\mu}^{ij}$ of 
Eq. (\ref{200}).

Let us comment on the 
relation between our result and that of other authors.
The second identity (\ref{4}) is also given in Ref. \cite{Xing2}.
The third identity (\ref{5}) is new and play an 
important role in deriving our exact formula.
The similar expression to (\ref{206}) is given in 
Ref. \cite{Ohlsson} 
as the result of the calculation of $e^{-i\tilde{H}L}$, 
although the CP phase has not been considered.
\footnote{We notice that the expression of $\tilde{U}\tilde{U}^*$ 
in (\ref{206}) is also derived from Eq. (4) in Ref. \cite{Harrison2}
after some calculations \cite{Harrison3}.}
Next, ${\rm Im}(pq^*)$ in (\ref{7}) are rewritten as 
\begin{eqnarray}
{\rm Im}(pq^*)=1/(2E)^3{\rm Im}(H_{e\mu}H_{\mu\tau}
H_{\tau e})=
\Delta_{12}\Delta_{23}\Delta_{31}J,\label{205}
\end{eqnarray}
from (\ref{30}) and (\ref{2}).
Naumov-Harrison-Scott identity is reproduced by  
substituting (\ref{205}) into (\ref{7}).

\section{Separation of CP odd/even Parts}
\hspace*{\parindent}
In this section, we give a concrete expression 
for the oscillation probability 
and then, we study the dependence of the oscillation probability on 
the CP phase.

First let us consider the constants $p$ and $q$.
We use the standard parametrization 
\begin{eqnarray}
U_{\alpha i}&=&\left(
\begin{array}{ccc}
c_{13}c_{12} & c_{13}s_{12} & s_{13}e^{-i\delta} \\
-c_{23}s_{12}-s_{23}s_{13}c_{12}e^{i\delta}
& c_{23}c_{12}-s_{23}s_{13}s_{12}e^{i\delta}
& s_{23}c_{13} \\
s_{23}s_{12}-c_{23}s_{13}c_{12}e^{i\delta}
& -s_{23}c_{12}-c_{23}s_{13}s_{12}e^{i\delta}
& c_{23}c_{13}
\end{array}
\right),
\end{eqnarray}
where 
$\sin \theta_{ij}=s_{ij}$,
$\cos \theta_{ij}=c_{ij}$.
In addition, as the neutrino oscillation probabilities 
do not depend on the mass itself, but the mass square differences, 
we take $m_1^2=0, m_2^2=\Delta_{21}$ and 
$m_3^2=\Delta_{31}$ without loss of generality.
So, $p$ and $q$ are given by 
\begin{eqnarray}
p=p_1 e^{-i\delta}+p_2, \hspace{0.5cm}
q=q_1 e^{-i\delta}+q_2,
\end{eqnarray}
where $p_i$ and $q_i$ are real numbers; 
\begin{eqnarray}
&&p_1=(\Delta_{31}-\Delta_{21}s_{12}^2)
s_{23}s_{13}c_{13}, \hspace{0.5cm}
p_2=\Delta_{21}s_{12}c_{12}c_{23}c_{13}, \label{32} \\
&&q_1=-\Delta_{31}\Delta_{21}c_{12}^2 s_{23}s_{13}c_{13},
\hspace{0.5cm}
q_2=-\Delta_{31}\Delta_{21}s_{12}c_{12}c_{23}c_{13}.
\label{33}
\end{eqnarray}
Then, we have 
\begin{eqnarray}
|p|^2&=&p_1^2+p_2^2+2p_1 p_2 \cos \delta, \\
|q|^2&=&q_1^2+q_2^2+2q_1 q_2 \cos \delta, \\
{\rm Re}(pq^*)&=&p_1 q_1+p_2 q_2
+(p_1 q_2+q_1 p_2)\cos \delta, \\
{\rm Im}(pq^*)&=&(p_2 q_1-p_1 q_2)\sin \delta.
\end{eqnarray}
Therefore, the oscillation probability can be written in the form 
\begin{eqnarray}
P(\nu_e \to \nu_{\mu})=A\cos \delta+B\sin \delta+C,
\label{8}
\end{eqnarray}
from (\ref{200})-(\ref{7}). 
Note that $A, B$ and $C$ are independent of $\delta$ 
and the oscillation probability is expressed only by 
the linear terms in $\cos \delta$ and $\sin \delta$ 
up to a constant as described below.
This is one of our main results.
Here 
\begin{eqnarray}
A&=&\sum_{(ij)}^{{\rm cyclic}}A_{ij}
\sin^2 \left(\frac{\tilde{\Delta}_{ij}L}{4E}\right),
\label{31}
\\
B&=&\sum_{(ij)}^{{\rm cyclic}}B^{\prime}
\sin \left(\frac{\tilde{\Delta}_{ij}L}{2E}\right),
\label{203} \\
C&=&\sum_{(ij)}^{{\rm cyclic}}C_{ij}
\sin^2 \left(\frac{\tilde{\Delta}_{ij}L}{4E}\right),
\end{eqnarray}
are expressed by the products of the oscillation part 
dependent on $L$ and $A_{ij}, B^{\prime}$ and $C_{ij}$.
And then, $A_{ij}, B^{\prime}$ and $C_{ij}$ are given by 
\begin{eqnarray}
A_{ij}&=&\frac{-4[2p_1 p_2 \lambda_i \lambda_j+2q_1 q_2
+(p_1q_2+q_1 p_2)(\lambda_i+\lambda_j)]}
{\tilde{\Delta}_{ij}\tilde{\Delta}_{12}
\tilde{\Delta}_{23}\tilde{\Delta}_{31}}, \label{34} \\
B^{\prime}&=&\frac{-2(p_2 q_1-p_1 q_2)}
{\tilde{\Delta}_{12}\tilde{\Delta}_{23}\tilde{\Delta}_{31}},
\label{35} \\
C_{ij}&=&\frac{-4[(p_1^2+p_2^2) \lambda_i \lambda_j
+(q_1^2+q_2^2)
+(p_1q_1+q_2 p_2)(\lambda_i+\lambda_j)]}
{\tilde{\Delta}_{ij}\tilde{\Delta}_{12}
\tilde{\Delta}_{23}\tilde{\Delta}_{31}}, 
\label{36}
\end{eqnarray}
as the function of the masses and mixing angles.
Since the effective masses $\lambda_i$ shown in 
\cite{Barger1, Zaglauer, Xing} do not depend on $\delta$, 
the coefficients $A, B$ and $C$ are independent of $\delta$. 

Our analytic result given in (\ref{8})
should be compared with the result of \cite{Minakata} depicted in Fig. 1
obtained numerically.
The trajectory becomes an ellipse in the
bi-probability space when $\delta$ changes from $0$ to $2\pi$.
The CP dependence of the exact form of $P(\nu_e \to \nu_{\mu})$
given in (\ref{8}) becomes much simpler than the result in \cite{Zaglauer}.
By solving (\ref{8}) for $\sin \delta$ and $\cos \delta$ 
one obtain  
\begin{eqnarray}
\sin \delta&=&\frac{B(P-C)
\pm A\sqrt{A^2+B^2-(P-C)^2}}
{A^2+B^2}, \label{201} \\
\cos \delta&=&\frac{A(P-C)
\mp B\sqrt{A^2+B^2-(P-C)^2}}
{A^2+B^2} \label{202}.
\end{eqnarray}
Thus, we can determine the value of CP phase except for the 
ambiguity of the sign  
from the measurement of the probability.
The sign ambiguity is understood as follows.
\begin{figure}
\begin{minipage}{7cm}
\begin{center}
\includegraphics[scale=0.7, bb=-40 460 282 700]{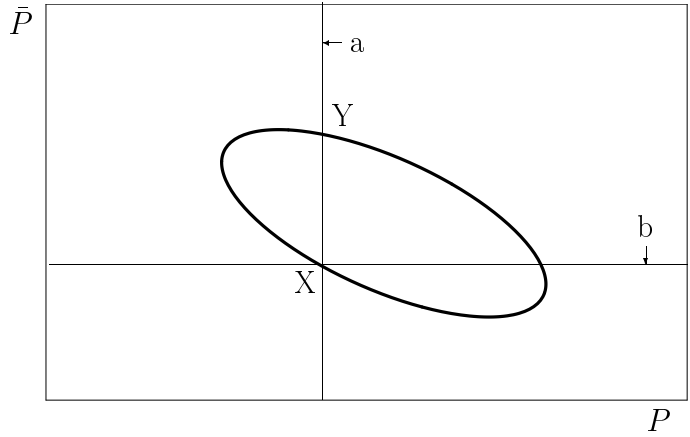}
\end{center}
\end{minipage}
\caption{An example of CP trajectory 
\hspace{0.5cm} We take $P$ for the horizontal axis and  
$\bar{P}$ for the vertical axis. The value of $\delta$ changes 
from $0$ to $2\pi$.}
\label{fig:1}       % Give a unique label
\end{figure}
If we measure the probability of the neutrino at a fixed energy 
and a baseline, we find the solutions on a ``line a".
As shown in Fig. 1, 
there are two intersections X and Y of ``line a" with 
the CP trajectory.
This is the reason why the ambiguity due to the sign 
appears in the analytic solutions (\ref{201}) and (\ref{202}).

In order to resolve the sign ambiguity, we need to measure 
more than two kinds of probabilities, for example, neutrino 
and antineutrino.
We denote $P$ and $\bar{P}$ of the oscillation probabilities for 
neutrino and antineutrino respectively as  
\begin{eqnarray}
P&=&A \cos \delta
+B \sin \delta+C, \\
\bar{P}&=&\bar{A} \cos \delta
+\bar{B} \sin \delta+\bar{C}.
\end{eqnarray}
Then, CP phase can be determined by   
\begin{eqnarray}
\sin \delta&=&\frac{(\bar{A} P-A \bar{P})
-(\bar{A} C-A \bar{C})}{\bar{A} B-A \bar{B}}, \label{100}\\
\cos \delta&=&\frac{(\bar{B} P-B \bar{P})
-(\bar{B} C-B \bar{C})}{\bar{B} A-B \bar{A}}, \label{101}
\end{eqnarray}
without the ambiguity of the sign.
This means that the solution is at X, the intersect of ``line a" and 
``line b".

Although the value of CP phase is determined, 
in principle, in (\ref{100}) and (\ref{101}), 
there remain other ambiguities 
included in $A$, $B$, $C$ and $\bar{A}$, $\bar{B}$, $\bar{C}$.
The methods to resolve these ambiguities are discussed 
in the references for example \cite{Minakata, Burguet, Barger2, Kajita}.
We discuss the ambiguities due to the sign of mass 
squared differences in Sec. 5.

\section{Simple Derivation of Approximate Formula}
\hspace*{\parindent}
In the previous section, we have shown that the exact formula of 
the oscillation probability can be expressed as 
$P(\nu_e \to \nu_{\mu})=A\cos \delta+B\sin \delta +C$.
In this section, we demonstrate that the approximate formula  
seen in \cite{Cervera, Freund} is easily derived as an example 
in the case of $m_1<m_2 \ll m_3$.
One obtains the approximate formula for other patterns of mass 
hierarchy in the same way.

Let us first consider the coefficient $B$ of $\sin \delta$.
$B$ is expressed in the form of the sum as (\ref{203}).
Note that, under the condition $x+y+z=0$, the identity  
\begin{eqnarray}
\sin 2x+\sin 2y +\sin 2z=-4\sin x \sin y \sin z,
\end{eqnarray}
holds, and $B$ from (\ref{203}) is rewritten 
in the form of product as
\begin{eqnarray}
B&=&\sum_{(ij)}^{{\rm cyclic}}B^{\prime}
\sin \left(\frac{\tilde{\Delta}_{ij}L}{2E}\right) \\
&=&-4B^{\prime}\sin \left(\frac{\tilde{\Delta}_{12}L}{4E}\right)
\sin \left(\frac{\tilde{\Delta}_{23}L}{4E}\right)
\sin \left(\frac{\tilde{\Delta}_{31}L}{4E}\right).
\end{eqnarray}
Next, let us consider the coefficient $A$ of $\cos \delta$.
Under the same condition as in deriving $B$, the identity 
\begin{eqnarray}
\sin^2 x=-(\sin x \sin y \cos z+\sin x \cos y \sin z).
\end{eqnarray}
holds and $A$ is rewritten as 
\begin{eqnarray}
A&=&\sum_{(ij)}^{{\rm cyclic}}A_{ij}\sin^2
\left(\frac{\tilde{\Delta}_{ij}L}{4E}\right) \\
&=&-\sum_{(ijk)}^{{\rm cyclic}}(A_{jk}+A_{ki})
\cos \left(\frac{\tilde{\Delta}_{ij}L}{4E}\right)
\sin \left(\frac{\tilde{\Delta}_{jk}L}{4E}\right)
\sin \left(\frac{\tilde{\Delta}_{ki}L}{4E}\right).
\end{eqnarray}

Substituting (\ref{32}) and (\ref{33}) for $p$ and $q$ in 
(\ref{34})-(\ref{36}), 
$A$, $B$ and $C$ are rewritten with the masses and the mixings as 
\begin{eqnarray}
A&=&\sum_{(ijk)}^{{\rm cyclic}}\frac{-8J_r \Delta_{21}
[\Delta_{31}\lambda_k(\lambda_k-\Delta_{31})
+A_k^{(1)}]}
{\tilde{\Delta}_{jk}^2\tilde{\Delta}_{ki}^2}
\cos \left(\frac{\tilde{\Delta}_{ij}L}{4E}\right)
\sin \left(\frac{\tilde{\Delta}_{jk}L}{4E}\right)
\sin \left(\frac{\tilde{\Delta}_{ki}L}{4E}\right),
\label{15} \\
B&=&\frac{8\Delta_{12} \Delta_{23} \Delta_{31}}
{\tilde{\Delta}_{12} \tilde{\Delta}_{23}
\tilde{\Delta}_{31}} J_r
\sin\left(\frac{\tilde{\Delta}_{12}L}{4E}\right)
\sin\left(\frac{\tilde{\Delta}_{23}L}{4E}\right)
\sin\left(\frac{\tilde{\Delta}_{31}L}{4E}\right), 
\label{16} \\
C&=&\sum_{(ij)}^{{\rm cyclic}}
\frac{-4[s_{13}^2 (s_{23}^2 c_{13}^2
\Delta_{31}^2 \lambda_i \lambda_j
+C_{ij}^{(1)}
+C_{ij}^{(2a)})+C_{ij}^{(2b)}]}
{\tilde{\Delta}_{ij} \tilde{\Delta}_{12}
\tilde{\Delta}_{23}\tilde{\Delta}_{31}}
\sin^2 \left(\frac{\tilde{\Delta}_{ij}L}{4E}\right),
\label{17}
\end{eqnarray}
where $J_r=s_{12}c_{12}s_{23}c_{23}s_{13}c_{13}^2$, and  
\begin{eqnarray}
A_k^{(1)}&=&\Delta_{21}\{\Delta_{31}\lambda_k
(c_{12}^2-s_{12}^2)+\lambda_k^2 s_{12}^2
-\Delta_{31}^2 c_{12}^2\}, \label{400} \\
C_{ij}^{(1)}&=&\Delta_{21}\Delta_{31}\{
-\lambda_i(\lambda_j s_{12}^2+\Delta_{31}c_{12}^2)
-\lambda_j(\lambda_i s_{12}^2+\Delta_{31}c_{12}^2)\}
s_{23}^2 c_{13}^2, \label{401}
\\
C_{ij}^{(2a)}&=&\Delta_{21}^2 
(\lambda_i s_{12}^2+\Delta_{31}c_{12}^2)
(\lambda_j s_{12}^2+\Delta_{31}c_{12}^2)
s_{23}^2 c_{13}^2, \label{402}
\\
C_{ij}^{(2b)}&=&\Delta_{21}^2 
(\lambda_i-\Delta_{31})(\lambda_j-\Delta_{31})
s_{12}^2 c_{12}^2 c_{23}^2 c_{13}^2. \label{403}
\end{eqnarray}
Note that these expressions are still exact.
In the limit of small $\Delta_{21}$, terms given in 
Eqs. (\ref{400})-(\ref{403}) are higher order in $\Delta_{21}$ 
and can be ignored.
The superscripts of $A$ and $C$ stand for the power of $\Delta_{21}$, 
and $(2a)$ represents the term proportional to $s_{13}^2$ and 
$(2b)$ is the term independent of $s_{13}^2$.

Finally, we obtain the well known approximate formula 
by neglecting the smallest effective mass.
In the high energy neutrino 
the smallest effective mass is $\lambda_1\simeq \Delta_{21}$. 
Other effective masses $\lambda_2$ and $\lambda_3$, 
correspond to $a$ or $\Delta_{31}$.
Accordingly, $A$, $B$ and $C$ are approximated by  
\begin{eqnarray}
A&=&\frac{8J_r \Delta_{21}\Delta_{31}}
{a(\Delta_{31}-a)}
\cos \left(\frac{\Delta_{31}L}{4E}\right)
\sin \left(\frac{aL}{4E}\right)
\sin \left(\frac{(\Delta_{31}-a)L}{4E}\right),
\label{90} \\
B&=&\frac{8J_r \Delta_{21} \Delta_{31}}
{a(\Delta_{31}-a)} 
\sin\left(\frac{\Delta_{31}L}{4E}\right)
\sin\left(\frac{aL}{4E}\right)
\sin\left(\frac{(\Delta_{31}-a)L}{4E}\right), 
\label{91} \\
C&=&\frac{4\Delta_{31}^2}
{(\Delta_{31}-a)^2}
s_{23}^2 s_{13}^2 c_{13}^2
\sin^2 \left(\frac{(\Delta_{31}-a)L}{4E}\right),
\label{92}
\end{eqnarray}
under the condition $\Delta_{21}/\Delta_{31} < s_{13}$.
When $s_{13}$ is smaller than $(\Delta_{21}/\Delta_{31})$, 
the term $C_{ij}^{(2b)}$ independent of $s_{13}$ 
becomes the dominant term.
Although the approximate formula derived here is in agreement with 
the ones seen in \cite{Cervera, Freund}, 
the derivation is rather simple.
Moreover, one can easily reproduce the approximate formula 
in low-energy \cite{Sato}.

\section{Numerical Analysis of CP odd/even Part}
\hspace*{\parindent}
In this section, we investigate the values of the coefficients 
$A, B$ and $C$ in cases of neutrino and antineutrino 
using the exact expressions.
We also investigate them changing 
the signs of $\Delta_{31}$ and $\Delta_{21}$.

In this numerical analysis, 
we take $\theta_{12}=\pi/4$, $|\Delta_{21}|=10^{-4} {\rm eV}^2$, 
$\theta_{23}=\pi/4$ and $|\Delta_{31}|=3 \times 10^{-3} {\rm eV}^2$ 
to be consistent with the LMA MSW solution to the solar neutrino 
problem \cite{solar, SNO} and the zenith-angle dependences of atmospheric 
neutrinos \cite{atmospheric}.
We also take $\theta_{13}=0.05$ 
within the upper bound of CHOOZ experiment \cite{CHOOZ}. 
The baseline length is typically taken to be  
$L=2900 {\rm km}$ and the matter density is taken to be 
$3.2{\rm g/cm}^3$. 

In Fig. 2 we show the coefficients $A$, $B$ and $C$ 
changing with the energy $E$. 
\psfrag{a}[][]{(a) $\nu_e \to \nu_{\mu}, m_1<m_2 \ll m_3$}
\psfrag{b}[][]{(b) $\bar{\nu}_e \to \bar{\nu}_{\mu}, m_1<m_2 \ll m_3$}
\psfrag{c}[][]{(c) $\nu_e \to \nu_{\mu}, m_3 \ll m_1<m_2$}
\psfrag{d}[][]{(d) $\bar{\nu}_e \to \bar{\nu}_{\mu}, m_3 \ll m_1<m_2$}
\psfrag{e}[][]{(e) $\nu_e \to \nu_{\mu}, m_2<m_1 \ll m_3$}
\psfrag{f}[][]{(f) $\bar{\nu}_e \to \bar{\nu}_{\mu}, m_2<m_1 \ll m_3$}
\psfrag{g}[][]{(g) $\nu_e \to \nu_{\mu}, m_3 \ll m_2 < m_1$}
\psfrag{h}[][]{(h) $\bar{\nu}_e \to \bar{\nu}_{\mu}, m_3 \ll m_2< m_1$}
\psfrag{egev}[][]{E (GeV)}
\begin{figure}
%\vspace*{0.5cm}
%\begin{minipage}{7cm}
\hspace{1cm}
\includegraphics[scale=1.1]{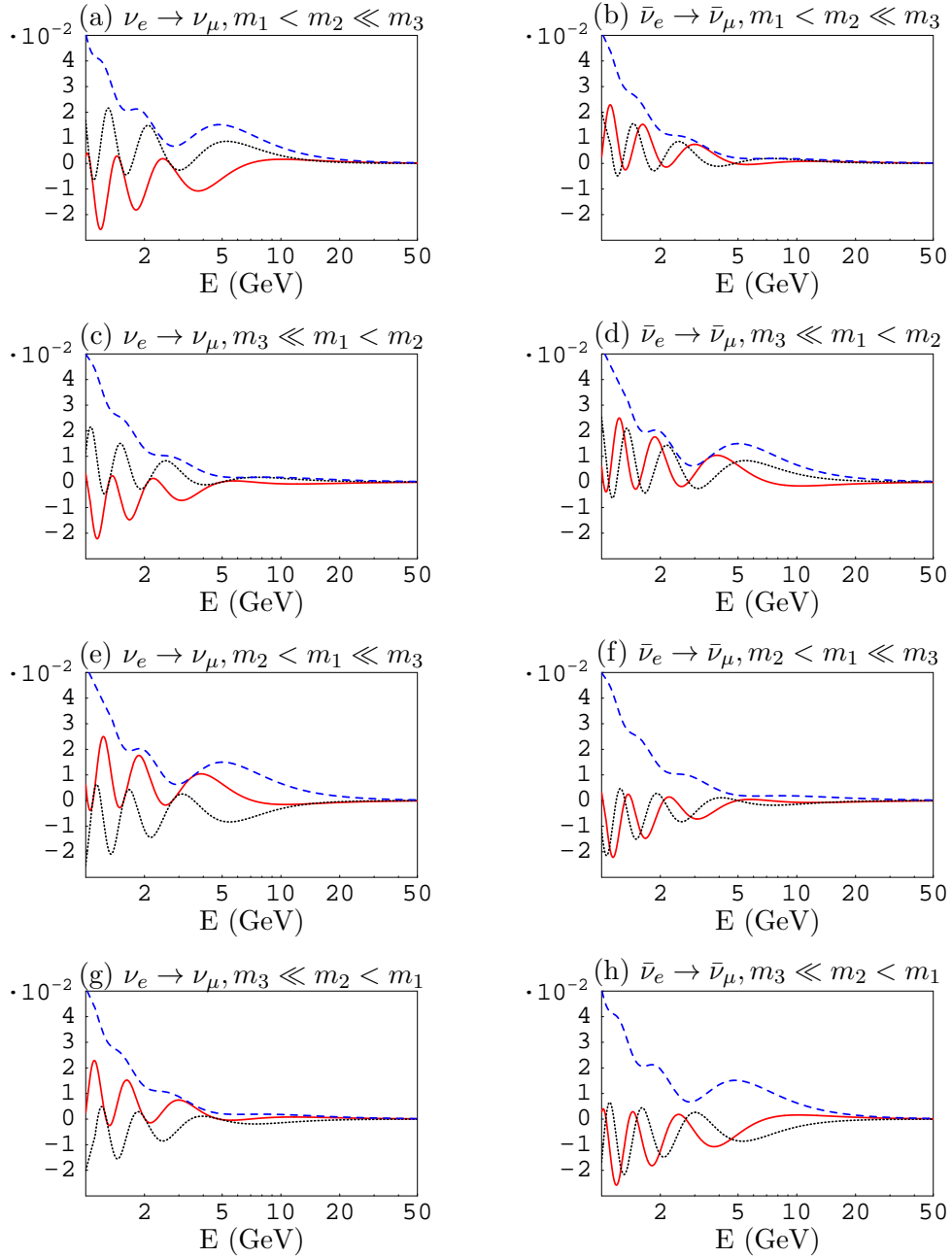}
%\end{minipage}
\caption{$A, B, C$ at $L=2900 \,{\rm km}$ \hspace{0.5cm}
The graphs of the left and right side correspond to the neutrino and the antineutrino 
respectively.
The solid lines, the dotted lines and the dashed lines are for $A$, 
$B$ and $C$ in all graphs.
And from top to bottom, 
$(\Delta_{31}>0, \Delta_{21}>0)$, $(\Delta_{31}<0, \Delta_{21}>0)$, 
$(\Delta_{31}>0, \Delta_{21}<0)$ and $(\Delta_{31}<0, \Delta_{21}<0)$ cases.}
\label{fig:2}       % Give a unique label
\end{figure}
We observe that the sign of $A$ is opposite for example
in Fig. 2(a) and (d).
We also observe that $A$ and $B$ have the opposite sign 
but $C$ has the same sign comparing Fig. 2(a) with (e).
In addition, some peaks have appeared in all graphs of Fig. 2 
with the change of energy.
In case of $\Delta_{31}>0$,
the peaks around $6 \,{\rm GeV}$ in Fig. 2(a) for neutrinos 
are enhanced compared with those in Fig. 2(b) for antineutrinos.
Inversely, in case of $\Delta_{31}<0$, 
the peaks in Fig. 2(d) for antineutrinos are enhanced 
compared with those in Fig. 2(c) for neutrinos.

These features are understood qualitatively from 
the approximate formula (\ref{90})-(\ref{92}).
First let us consider the sign of $A$, $B$ and $C$.
As we found from (\ref{90})-(\ref{92}), when the signs of 
both $\Delta_{31}$ and $a$ change, the sign of $A$ becomes 
opposite and the signs of $B$ and $C$ do not change.
On the other hand, when the sign of $\Delta_{21}$ changes, 
the signs of both $A$ and $B$ change while the sign of $C$ 
does not change.
Next, let us consider the magnitude of the peak around $6 \,{\rm GeV}$.
These are strongly affected by the denominator $(\Delta_{31}-a)$. 
Since the signs of $\Delta_{31}$ and $a$ are opposite
in Fig. 2(a) and (d),
the denominator $(\Delta_{31}-a)$ becomes small and 
the magnitude of the peaks are enhanced.
On the other hand, 
since the signs of $\Delta_{31}$ and $a$ are the same
in Fig. 2(b) and (c), the peaks are suppressed.
Finally, let us explain the position of the peak in Fig. 2(a) and (d) 
around $6\, {\rm GeV}$.
Roughly, the peak energy is determined by the following:  
\begin{eqnarray}
\sin \left[1.27\left(\frac{\Delta_{31}-a}{1 \,{\rm eV}^2}\right)
\left(\frac{L}{1 \,{\rm km}}\right)
\left(\frac{E}{1 \,{\rm GeV}}\right)^{-1}\right] \sim 1 
\to E \simeq 6 \,{\rm GeV} 
\hspace{0.5cm}{\rm (at} \,\, L=2900 \,{\rm km)}. 
\end{eqnarray}
As pointed out by Parke and Weiler \cite{Parke}, and Lipari 
\cite{Lipari}, the peak energy is lower than the energy 
of 1-3 MSW resonance since the baseline length is short 
compared with the earth diameter.
The above discussions on Fig. 2(a)-(e) 
can be applied to other figures.

We have studied how the magnitude of $A$, $B$ and $C$ 
change due to the sign of the mass squared differences.
In the case of $m_1<m_2 \ll m_3$, the coefficients have been 
investigated in Ref. \cite{Cervera} by using the approximate formula.
These correspond to Fig. 2(a) and (b).
The sign of $\Delta_{31}$ is determined from the leading term $C$ 
as pointed out by many authors (for example see \cite{Cervera}). 
On the other hand, the sign of $\Delta_{21}$ is determined from 
next leading terms $A$ or $B$.
This means that the sign of $\Delta_{21}$ is simultaneously determined 
in addition to the CP phase.
It may be interesting as the first observation of 
the sign of $\Delta_{21}$ using artificial neutrino beam.

\section{Summary}
\hspace*{\parindent}
We have studied neutrino oscillations in constant matter 
within the framework of the three neutrino scenario.
We summarize the results obtained in this letter.

\begin{enumerate}
\renewcommand{\labelenumi}{(\roman{enumi})}

\item  We have derived an exact expression of the oscillation 
probability by using a new method.
We have calculated $\tilde{U}\tilde{U}^*$ from the identities  
without directly calculating single $\tilde{U}$.
Not only the derivation but also the result becomes simple 
and the matter effects enter only through effective masses.

\item  We have obtained the CP dependence of the oscillation 
probability exactly by using the standard parametrization.
It has been shown that the oscillation probability is in the form,  
$P(\nu_e \to \nu_{\mu})=A\cos \delta+B\sin \delta+C$.
We have also demonstrated that the approximate formula in 
high-energy can be easily reproduced from our result.
  
\end{enumerate}

Finally, let us comment on the oscillation probabilities 
for other channels.
These probabilities are easily derived 
in the same way as $P(\nu_e \to \nu_{\mu})$.
$P(\nu_e \to \nu_{\tau})$ has the same CP dependence as 
$P(\nu_e \to \nu_{\mu})$.
However, $P(\nu_{\mu} \to \nu_{\tau})$ has the term 
which depends on $\cos 2\delta$ in addition to the linear 
terms in $\sin \delta$ and $\cos \delta$.

\vspace{20pt}
\noindent
{\Large {\bf Acknowledgement}}

\noindent
The authors would like to thank Prof. A. I. Sanda 
for reading the manuscript and making a number of 
helpful suggestions.
We would like to thank Prof. H. Minakata and Prof. O. Yasuda 
for discussions and valuable comments.

\end{document}